\documentclass[11pt]{article}
\usepackage[margin=1in]{geometry}
\usepackage{amsmath,amssymb,mathtools}
\usepackage{graphicx}
\usepackage{hyperref}

\title{From Ringdown to Lensing: Analytic Eikonal Modes of Quasi-Topological Regular Black Holes}
\author{Alexey Dubinsky\\
\small University of Seville, Seville, Spain\\
\small \texttt{dubinsky@ukr.net}}
\date{}

\begin{document}
\maketitle

\begin{abstract}
We develop an analytic eikonal description of perturbations for four-dimensional regular black holes in quasi-topological gravity. Using first-order Schutz--Will WKB together with a small-coupling expansion and a large-$\ell$ expansion, we obtain closed quasinormal-mode formulas with explicit dependence on the black-hole parameters $(M,\mu,\nu,\alpha)$. We then map the same geodesic invariants $(\Omega_{\text{ph}},\lambda_{\text{ph}})$ to shadow and strong-lensing observables, deriving an explicit QNM--shadow--lensing correspondence. In this way, ringdown frequencies, shadow scale, and strong-deflection observables are unified in one analytic scheme for this quasi-topological family.
\end{abstract}

\section{Introduction}

Regular black holes in quasi-topological gravity are part of a broader higher-curvature program. Foundational constructions include cubic quasi-topological black holes and holographic applications in Refs.~\cite{MyersRobinson2010,MyersPaulosSinha2010}, together with the closely related cubic sector of Ref.~\cite{OlivaRay2010}. Higher-order quasi-topological extensions were developed at quartic and quintic order in Refs.~\cite{AhmedEtAl2017,CisternaEtAl2017}, while the generalized quasi-topological framework and its systematic all-order organization were established in Refs.~\cite{HennigarKubiznakMann2017,BuenoEtAl2019,BuenoCanoHennigar2020,BuenoEtAl2023Shebang}. Additional matter-coupled extensions were analyzed in Ref.~\cite{CanoMurcia2020}.

Within this landscape, recent four-dimensional regular-black-hole realizations from pure gravity were presented in Refs.~\cite{Bueno2025,BorissovaCarballo2026}, and non-singular cosmological implications in $d=4$ non-polynomial quasi-topological gravity were discussed in Ref.~\cite{BorissovaMagueijo2026}. For the specific quasi-topological regular family studied here, Ref.~\cite{Konoplya2026} analyzed quasinormal spectra (including overtone behavior), while Ref.~\cite{Malik2026} focused on orbit dynamics and optical signatures, including black-hole shadows and related geodesic observables. A complementary line of work by Malik developed semi-analytic and beyond-eikonal quasinormal-mode treatments across different black-hole backgrounds: bumblebee black holes with a global monopole \cite{Malik2023Bumblebee}, Dirac-field perturbations in effective quantum-gravity black holes \cite{Malik2024DiracEQG}, massive-field QNM/grey-body correspondence in Schwarzschild--de Sitter spacetime \cite{Malik2024SdSMassive}, long-lived modes for brane-localized Reissner--Nordstr\"om--de Sitter black holes \cite{Malik2025BraneRNdS}, and analytic beyond-eikonal formulas for a regular Dehnen-halo black hole \cite{Malik2026Dehnen}. On the methodological side, Zhidenko and collaborators derived widely used semi-analytic formulas for quasinormal spectra in the eikonal and near-eikonal regimes, from the early asymptotic analysis of Schwarzschild--de Sitter quasinormal frequencies \cite{Zhidenko2004SdS} to higher-order WKB/eikonal expansions and compact analytic expressions for QNMs and grey-body factors \cite{Konoplya2003WKB,KonoplyaZhidenkoZinhailo2019,KonoplyaZhidenko2023Eikonal,Dubinsky:2026wcv}, as well as analytic near-extremal de Sitter formulas in Kerr--Newman--de Sitter backgrounds \cite{ChurilovaKonoplyaZhidenko2022}. Related eikonal-focused studies include large-$\ell$ analytic behavior in Schwarzschild--G\"odel backgrounds \cite{KonoplyaAbdalla2005Godel}, eikonal-instability maps for Einstein--Gauss--Bonnet and Lovelock black holes \cite{KonoplyaZhidenko2017GBAdS,KonoplyaZhidenko2017Lovelock}, and asymptotically safe regular-black-hole ringdown analyses where eikonal behavior is contrasted with overtone effects \cite{KonoplyaEtAl2022ASG}.

At the same time, the geometric correspondence between eikonal quasinormal modes and unstable null geodesics is now standard in general relativity \cite{Cardoso2009}, and it can be extended to strong gravitational lensing observables in the Stefanov--Yazadjiev--Gyulchev program \cite{Stefanov2010}. However, this correspondence is not always exact beyond test-field/eikonal conditions, and several subtleties are known. In particular, limits and failures of the naive geodesic/QNM duality for gravitational perturbations were clarified in Refs.~\cite{KonoplyaStuchlik2017,Konoplya2022Duality}, and a concrete recent example of breakdown in higher-curvature gravity was shown for Starobinsky--Bel--Robinson black holes in Ref.~\cite{Bolokhov2023}. Therefore, for the quasi-topological regular-black-hole family, it is important to separate what can be derived analytically and robustly in the eikonal sector from what requires full numerical perturbation theory.

The aim of this paper is to provide that analytic sector in closed form. Starting from the explicit metric function, we derive eikonal QNMs through first-order Schutz--Will WKB and a small-coupling expansion, keeping explicit dependence on $(M,\mu,\nu,\alpha)$. We then show how the same invariants $\Omega_{\text{ph}}$ and $\lambda_{\text{ph}}$ govern the shadow radius and strong-deflection lensing coefficients. To the best of our knowledge, this yields the first explicit analytic eikonal formulas for this quasi-topological metric family and the first direct QNM--shadow--lensing map for it.

The paper is organized as follows. In Section~2 we present the quasi-topological geometry and perturbative setup; in Section~3 we derive photon-sphere quantities in the small-coupling regime; in Section~4 we obtain the first-order WKB eikonal quasinormal frequencies; in Section~5 we establish the shadow--ringdown correspondence; in Section~6 we connect the same invariants to strong-lensing observables and the Stefanov-type map; and in Section~7 we summarize the main results and implications.

\section{Geometry and perturbative setup}

Following Ref.~\cite{Konoplya2026}, we consider the static spherical geometry
\begin{equation}
ds^2=-f(r)dt^2+\frac{dr^2}{f(r)}+r^2(d\theta^2+\sin^2\theta\,d\varphi^2),
\label{eq:metric}
\end{equation}
with
\begin{equation}
f(r,\alpha)=1-r^2\psi(r,\alpha).
\label{eq:fpsi}
\end{equation}
The explicit expression for $\psi$ is
\begin{equation}
\psi(r,\alpha)=\frac{2M\,r^{\mu-3}}{\left(r^{\nu}+\alpha^{\nu/3}(2M)^{\nu/3}\right)^{\mu/\nu}},
\label{eq:psi-explicit}
\end{equation}
which gives
\begin{equation}
f(r,\alpha)=1-\frac{2M\,r^{\mu-1}}{\left(r^{\nu}+\alpha^{\nu/3}(2M)^{\nu/3}\right)^{\mu/\nu}}.
\label{eq:fexplicit}
\end{equation}
It is convenient to define the deformation scale
\begin{equation}
\varepsilon\equiv \alpha^{\nu/3}(2M)^{\nu/3},
\qquad \varepsilon\ll r^{\nu},
\label{eq:epsilondef}
\end{equation}
so that
\begin{equation}
f(r,\alpha)=1-\frac{2M}{r}\left(1+\frac{\varepsilon}{r^{\nu}}\right)^{-\mu/\nu}.
\end{equation}
Expanding to first non-vanishing order in $\alpha$ (equivalently first order in $\varepsilon$) gives
\begin{equation}
f(r,\alpha)=f_0(r)+\varepsilon\,p(r)+\mathcal{O}(\varepsilon^2),
\quad
f_0(r)=1-\frac{2M}{r},
\quad
 p(r)=\frac{2M\mu}{\nu\,r^{\nu+1}}.
\label{eq:fexpand}
\end{equation}
The eikonal expansion is organized with the Langer combination
\begin{equation}
L\equiv \ell+\frac12,
\end{equation}
and $L\gg1$.

\section{Photon sphere data in the small-alpha regime}

Black-hole shadow observables provide a direct probe of near-horizon null geodesics and have become a central tool in strong-gravity phenomenology, from early theoretical imaging proposals to precision shape diagnostics and horizon-scale observations \cite{FalckeMeliaAgol2000,HiokiMaeda2009,JohannsenPsaltis2010,EHTM87VI2019}. For parametrized rotating geometries, practical shadow constructions and their analytic interpretation were developed in detail in Refs.~\cite{KonoplyaZhidenko2021Shadows,PerlickTsupko2022Review}, which motivates our explicit photon-sphere expansion below.

For null geodesics, the effective potential is $V_{\text{null}}(r)=L_z^2f(r)/r^2$. The unstable circular orbit radius $r_{\text{ph}}$ solves
\begin{equation}
r f'(r)-2f(r)=0.
\label{eq:phconds}
\end{equation}
To perform the small-$\alpha$ calculation transparently, we define
\begin{equation}
\mathcal{P}(r,\alpha)\equiv r f'(r,\alpha)-2f(r,\alpha)
=\mathcal{P}_0(r)+\varepsilon\mathcal{P}_1(r)+\mathcal{O}(\varepsilon^2),
\end{equation}
with
\begin{equation}
\mathcal{P}_0(r)=r f_0'(r)-2f_0(r)=\frac{6M}{r}-2,
\qquad
\mathcal{P}_1(r)=r p'(r)-2p(r).
\label{eq:Pparts}
\end{equation}
Now expand
\begin{equation}
r_{\text{ph}}=3M+\varepsilon\,\rho_1+\mathcal{O}(\varepsilon^2),
\label{eq:rphansatz}
\end{equation}
and impose $\mathcal{P}(r_{\text{ph}},\alpha)=0$. Since $\mathcal{P}_0(3M)=0$, the first nontrivial order gives
\begin{equation}
\rho_1\,\mathcal{P}_0'(3M)+\mathcal{P}_1(3M)=0,
\qquad
\mathcal{P}_0'(3M)=-\frac{2}{3M}.
\end{equation}
Hence
\begin{equation}
\rho_1=\frac{3M}{2}\left[3M p'(3M)-2p(3M)\right]
=-\frac{\mu(\nu+3)}{\nu\,3^{\nu}M^{\nu-1}},
\end{equation}
that is,
\begin{equation}
r_{\text{ph}}=3M-\varepsilon\frac{\mu(\nu+3)}{\nu\,3^{\nu}M^{\nu-1}}+\mathcal{O}(\varepsilon^2).
\label{eq:drph}
\end{equation}

The angular frequency and Lyapunov exponent of the circular null orbit are
\begin{equation}
\Omega_{\text{ph}}=\frac{\sqrt{f(r_{\text{ph}})}}{r_{\text{ph}}},
\qquad
\lambda_{\text{ph}}=\sqrt{\frac{f(r_{\text{ph}})\left[2f(r_{\text{ph}})-r_{\text{ph}}^2f''(r_{\text{ph}})\right]}{2r_{\text{ph}}^2}},
\label{eq:omegalambda}
\end{equation}
and we now expand each explicitly.

For $\Omega_{\text{ph}}$, write
\begin{equation}
\Omega^2_{\text{ph}}=\frac{f(r_{\text{ph}})}{r_{\text{ph}}^2}
=g(r_{\text{ph}}),
\qquad
g(r)\equiv \frac{f(r)}{r^2}.
\end{equation}
At zeroth order, $g_0'(3M)=0$ (equivalent to the photon-sphere condition), so the linear correction from $\rho_1$ cancels. Therefore
\begin{equation}
\Omega^2_{\text{ph}}=\frac{1}{27M^2}+\varepsilon\,\frac{p(3M)}{9M^2}+\mathcal{O}(\varepsilon^2),
\end{equation}
which yields
\begin{equation}
\Omega_{\text{ph}}=\frac{1}{3\sqrt{3}M}\left[1+\frac{3\varepsilon}{2}p(3M)\right]+\mathcal{O}(\varepsilon^2)
=\frac{1}{3\sqrt{3}M}\left[1+\frac{\mu\,\varepsilon}{\nu\,3^{\nu}M^{\nu}}\right]+\mathcal{O}(\varepsilon^2),
\label{eq:omegaalpha}
\end{equation}

For $\lambda_{\text{ph}}$, define
\begin{equation}
\Xi(r)\equiv \frac{f(r)\left[2f(r)-r^2f''(r)\right]}{2r^2},
\qquad
\lambda_{\text{ph}}=\sqrt{\Xi(r_{\text{ph}})}.
\end{equation}
Expanding around $\alpha=0$ and $r=3M$ gives
\begin{equation}
\Xi(r_{\text{ph}})=\frac{1}{27M^2}
\left[1+2\varepsilon\left(2p(3M)-\frac{9M^2}{4}p''(3M)\right)\right]
+\mathcal{O}(\varepsilon^2),
\end{equation}
so that
\begin{equation}
\lambda_{\text{ph}}=\frac{1}{3\sqrt{3}M}
\left[1-\frac{\mu(\nu^2+3\nu-6)}{2\nu\,3^{\nu+1}M^{\nu}}\,\varepsilon\right]
+\mathcal{O}(\varepsilon^2).
\label{eq:lambdaalpha}
\end{equation}

\begin{figure}[t]
\centering
\includegraphics[width=0.98\linewidth]{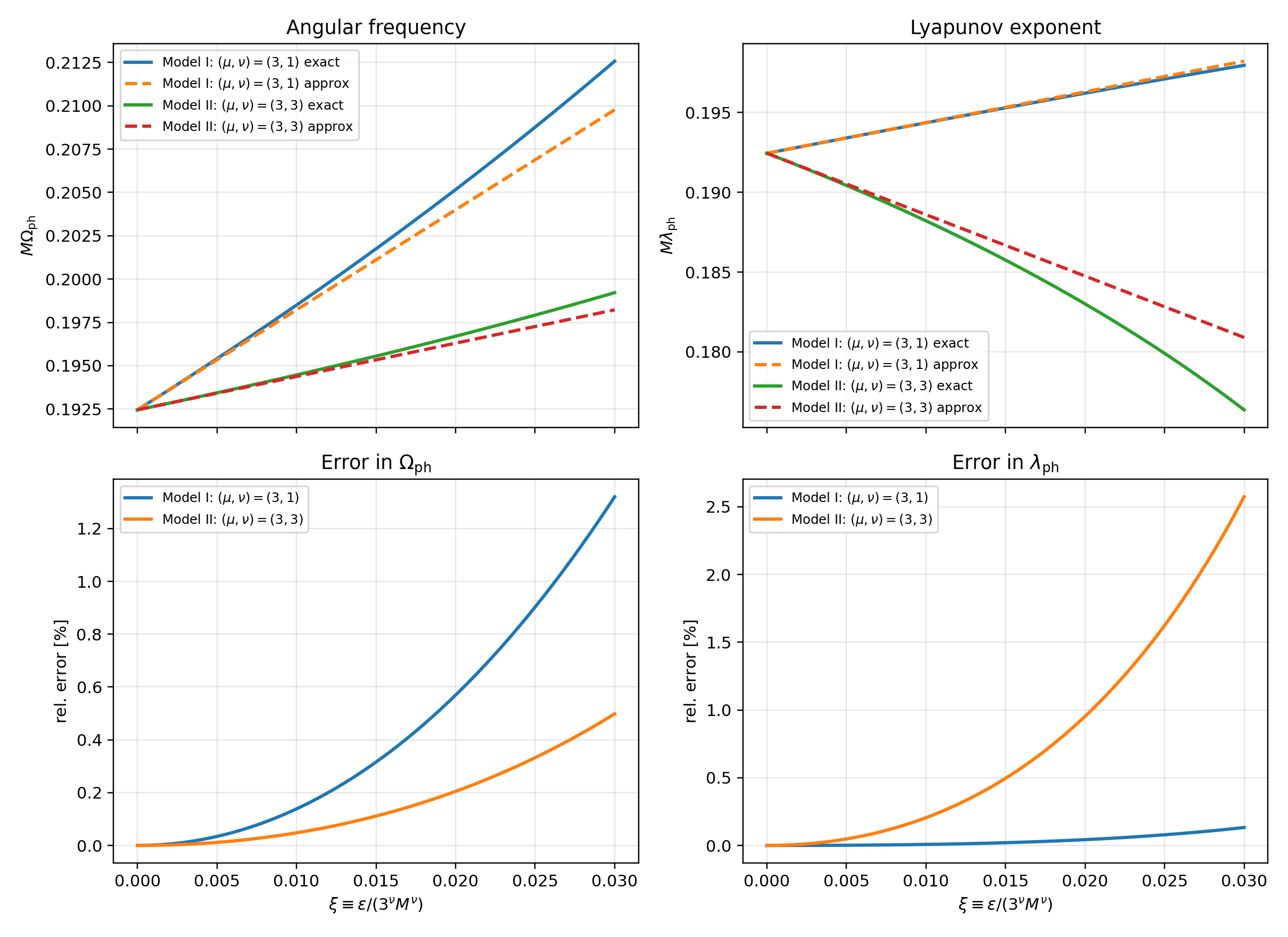}
\caption{Exact versus first-order small-$\varepsilon$ approximations for $\Omega_{\text{ph}}$ and $\lambda_{\text{ph}}$ using the full metric of Eq.~\eqref{eq:fexplicit}. Top panels: $M\Omega_{\text{ph}}$ and $M\lambda_{\text{ph}}$ as functions of $\xi\equiv\varepsilon/(3^{\nu}M^{\nu})$ for Models I and II. Bottom panels: corresponding relative errors (in percent).}
\label{fig:omega-lambda-check}
\end{figure}

To compare these first-order formulas with the exact (non-expanded) values, we solve Eq.~\eqref{eq:phconds} numerically using the full metric function in Eq.~\eqref{eq:fexplicit}, then evaluate $\Omega_{\text{ph}}$ and $\lambda_{\text{ph}}$ from Eq.~\eqref{eq:omegalambda}. Figure~\ref{fig:omega-lambda-check} shows this comparison for the two representative cases of Ref.~\cite{Konoplya2026}: Model I, $(\mu,\nu)=(3,1)$, and Model II, $(\mu,\nu)=(3,3)$.

We parametrize the small-coupling regime by
\begin{equation}
\xi\equiv\frac{\varepsilon}{3^{\nu}M^{\nu}},
\qquad
0\le \xi\le 0.03,
\end{equation}
which corresponds to $\varepsilon/r_{\text{ph}}^{\nu}\ll1$ near the photon sphere. Over this interval, the first-order approximation remains accurate: at $\xi=0.03$ we obtain relative errors
\begin{equation}
\delta\Omega\equiv\frac{|\Omega_{\text{approx}}-\Omega_{\text{exact}}|}{\Omega_{\text{exact}}}
=\begin{cases}
1.32\% & \text{Model I},\\
0.50\% & \text{Model II},
\end{cases}
\end{equation}
and
\begin{equation}
\delta\lambda\equiv\frac{|\lambda_{\text{approx}}-\lambda_{\text{exact}}|}{\lambda_{\text{exact}}}
=\begin{cases}
0.13\% & \text{Model I},\\
2.57\% & \text{Model II}.
\end{cases}
\end{equation}
Thus, in the small-$\xi$ domain relevant for the perturbative expansion, Eqs.~\eqref{eq:omegaalpha} and \eqref{eq:lambdaalpha} provide a quantitatively reliable approximation, with the $\lambda_{\text{ph}}$ correction in Model II showing the largest sensitivity to higher-order terms.

\section{First-order WKB and eikonal quasinormal modes}

The WKB approach is central for quasinormal-mode calculations because it provides a fast and physically transparent bridge between effective-potential barrier properties and complex frequencies, especially in the eikonal and moderate-overtone regime. Starting from the foundational Schutz--Will/Iyer--Will construction and higher-order extensions \cite{SchutzWill1985,IyerWill1987,Konoplya2003WKB,Matyjasek:2017psv,Konoplya:2026fqh}, a broad arXiv literature has applied and refined WKB in concrete black-hole backgrounds \cite{Malik:2026lfj,Konoplya:2001ji,Guo:2020caw,Kokkotas:2010zd,Abdalla:2005hu,Konoplya:2009hv,Pathrikar:2025gzu,Gonzalez:2022ote,Malik:2025erb,Fernando:2016ftj,Momennia:2018hsm,Bolokhov:2025lnt,Eniceicu:2019npi,Konoplya:2005sy,Bolokhov:2025aqy}, including Schwarzschild--de Sitter, Einstein--Aether, and improved semianalytic Pad\'e formulations \cite{KonoplyaZhidenko2004SdSHighOvertones,KonoplyaZhidenko2007EAether,MatyjasekOpala2017,KonoplyaZhidenkoZinhailo2019}. Recent examples by Bolokhov include WKB analyses in higher-curvature and quantum-corrected settings \cite{Bolokhov2023,Bolokhov2024Holonomy,Bolokhov2024EulerHeisenberg,Bolokhov2025EffectiveQG}; by Skvortsova in lower-dimensional, quantum-corrected and near-extreme compact-object backgrounds \cite{Skvortsova2024Spectrum2p1,Skvortsova2024QOS,Skvortsova2025RegularExtreme,Skvortsova2025QuantumCorrected}; and by L\"utf\"uo\u{g}lu in Weyl, Einstein--Yang--Mills and asymptotically safe gravity applications \cite{Lutfuoglu2025Weyl,Lutfuoglu2025EYM,Lutfuoglu2025ProperTime,Lutfuoglu2026ASG}.

For master perturbations written as
\begin{equation}
\frac{d^2\Psi}{dr_*^2}+\left[\omega^2-V_{\ell}(r)\right]\Psi=0,
\qquad \frac{dr_*}{dr}=\frac{1}{f(r)},
\end{equation}
the large-$L$ potential takes the form
\begin{equation}
V_{\ell}(r)=L^2\,\frac{f(r)}{r^2}+U(r)+\mathcal{O}(L^{-2}),
\label{eq:Vexpansion}
\end{equation}
where $U(r)$ is spin dependent (for instance $U=f f'/r$ for a scalar field). The first-order Schutz--Will condition \cite{SchutzWill1985,IyerWill1987}
\begin{equation}
\frac{iQ_0}{\sqrt{2Q_0''}}=n+\frac12,
\qquad Q(r)\equiv \omega^2-V_{\ell}(r),
\label{eq:SW}
\end{equation}
evaluated at the potential maximum $r_0$ in tortoise coordinate $r_*$ (primes in $Q_0''$ denote $d/dr_*$), with
\begin{equation}
r_0=r_{\text{ph}}+\mathcal{O}(L^{-2}),
\end{equation}
because the dominant term in Eq.~\eqref{eq:Vexpansion} is $L^2 f/r^2$.

At the peak, we have
\begin{equation}
V_0=L^2\frac{f_c}{r_c^2}+U_c+\mathcal{O}(L^{-2}),
\qquad
V_0''=L^2f_c^2\,g_c''+\mathcal{O}(1),
\end{equation}
where $r_c\equiv r_{\text{ph}}$, $f_c\equiv f(r_c)$, and $g(r)=f/r^2$. Using $r_c f_c'-2f_c=0$, one gets
\begin{equation}
g_c''=-\frac{2f_c-r_c^2f_c''}{r_c^4}<0,
\end{equation}
so that
\begin{equation}
\sqrt{-2V_0''}=2L\Omega_{\text{ph}}\lambda_{\text{ph}}+\mathcal{O}(L^{-1}),
\end{equation}
with $\Omega_{\text{ph}}$ and $\lambda_{\text{ph}}$ given by Eq.~\eqref{eq:omegalambda}. Inserting
\begin{equation}
\omega= L\Omega_{\text{ph}}+\omega^{(0)}+\mathcal{O}(L^{-1}),
\end{equation}
into Eq.~\eqref{eq:SW}, and separating real and imaginary parts order by order, gives
\begin{equation}
\operatorname{Re}\,\omega^{(0)}=\frac{U(r_{\text{ph}})}{2L\Omega_{\text{ph}}},
\qquad
\operatorname{Im}\,\omega^{(0)}=-\left(n+\frac12\right)\lambda_{\text{ph}},
\end{equation}
hence
\begin{equation}
\omega_{\ell n}=L\Omega_{\text{ph}}-i\left(n+\frac12\right)\lambda_{\text{ph}}+\frac{U(r_{\text{ph}})}{2L\Omega_{\text{ph}}}+\mathcal{O}(L^{-2}).
\label{eq:masterqnm}
\end{equation}

Substituting Eqs.~\eqref{eq:omegaalpha} and \eqref{eq:lambdaalpha} into Eq.~\eqref{eq:masterqnm}, the doubly expanded spectrum is
\begin{align}
\omega_{\ell n}
&=\frac{L}{3\sqrt{3}M}\left[1+\frac{\mu\,\varepsilon}{\nu\,3^{\nu}M^{\nu}}\right]
-i\frac{\left(n+\frac12\right)}{3\sqrt{3}M}
\left[1-\frac{\mu(\nu^2+3\nu-6)}{2\nu\,3^{\nu+1}M^{\nu}}\,\varepsilon\right]
\nonumber\\
&\quad +\frac{U(r_{\text{ph}})}{2L\Omega_{\text{ph}}}
+\mathcal{O}(\varepsilon^2,L^{-2},\varepsilon L^{-2}).
\label{eq:finalqnm}
\end{align}
Equation~\eqref{eq:finalqnm} is now explicit in the black-hole parameters $(M,\mu,\nu,\alpha)$ through Eq.~\eqref{eq:epsilondef}.
For the two representative cases emphasized in Ref.~\cite{Konoplya2026}, one gets: (i) Model I, $(\mu,\nu)=(3,1)$,
\begin{equation}
\Omega_{\text{ph}}=\frac{1}{3\sqrt{3}M}\left(1+\frac{\varepsilon}{M}\right),
\qquad
\lambda_{\text{ph}}=\frac{1}{3\sqrt{3}M}\left(1+\frac{\varepsilon}{3M}\right);
\end{equation}
and (ii) Model II, $(\mu,\nu)=(3,3)$, where $\varepsilon=2M\alpha$,
\begin{equation}
\Omega_{\text{ph}}=\frac{1}{3\sqrt{3}M}\left(1+\frac{2\alpha}{27M^2}\right),
\qquad
\lambda_{\text{ph}}=\frac{1}{3\sqrt{3}M}\left(1-\frac{4\alpha}{27M^2}\right).
\end{equation}

To make the spectral shift fully explicit for the models under consideration, we combine the eikonal frequency and the Lyapunov exponent into a single expression for $M\omega_{\ell n}$, yielding the following explicit results:
\begin{itemize}
    \item \textbf{Model I} $(\mu=3, \nu=1)$:
    \begin{equation}
        M\omega_{\ell n} \approx \frac{1}{3\sqrt{3}} \left[ L\left(1+\frac{\varepsilon}{M}\right) - i \left(n+\frac{1}{2}\right) \left(1+\frac{\varepsilon}{3M}\right) \right],
    \label{eq:qnm_mod1}
    \end{equation}
    \item \textbf{Model II} $(\mu=3, \nu=3)$ with $\varepsilon = 2M\alpha$:
    \begin{equation}
        M\omega_{\ell n} \approx \frac{1}{3\sqrt{3}} \left[ L\left(1+\frac{2\alpha}{27M^2}\right) - i \left(n+\frac{1}{2}\right) \left(1-\frac{4\alpha}{27M^2}\right) \right].
    \label{eq:qnm_mod2}
    \end{equation}
\end{itemize}
These expressions clearly show that while both models increase the oscillation frequency (real part), they exert opposite effects on the damping rate (imaginary part).

\begin{figure}[htbp]
    \centering
    \includegraphics[width=0.5\textwidth]{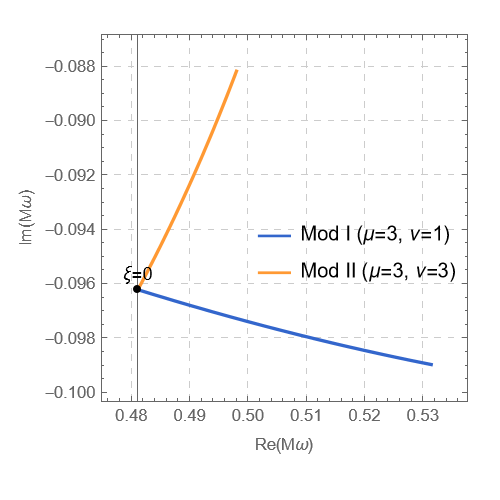} 
    \caption{The QNM spectrum map for the $\ell=2, n=0$ mode in the complex plane. The trajectories show the shift from the Schwarzschild point (black dot) as the coupling parameter $\xi$ varies from $0$ to $0.03$. Blue and orange lines correspond to Model I ($\mu=3, \nu=1$) and Model II ($\mu=3, \nu=3$), respectively.}
    \label{fig:qnm_map}
\end{figure}

The numerical trajectory of the fundamental quasinormal mode in the complex plane as a function of the deformation parameter $\xi$ is presented in Fig.~\ref{fig:qnm_map}. The black dot represents the Schwarzschild limit ($\xi = 0$). As $\xi$ increases, we observe a clear migration of the frequencies. For both models, the real part $\text{Re}(M\omega)$ increases, suggesting a higher oscillation frequency for these regular black hole solutions compared to the GR case. However, the damping behavior is model-dependent: while Model II ($\nu=3$) moves towards the real axis (lower damping), Model I ($\nu=1$) shows an increase in the absolute value of the imaginary part, indicating faster decay of the perturbations.

\section{Shadow--ringdown correspondence}

For a static observer at infinity, the shadow radius equals the critical impact parameter of the unstable photon orbit,
\begin{equation}
R_{\text{sh}}=b_c=\frac{r_{\text{ph}}}{\sqrt{f(r_{\text{ph}})}}=\frac{1}{\Omega_{\text{ph}}}.
\label{eq:shadowdef}
\end{equation}
The equality $R_{\text{sh}}=1/\Omega_{\text{ph}}$ is exact for any static spherical metric. Using Eq.~\eqref{eq:omegaalpha},
\begin{equation}
R_{\text{sh}}=3\sqrt{3}M\left[1-\frac{\mu\,\varepsilon}{\nu\,3^{\nu}M^{\nu}}\right]+\mathcal{O}(\varepsilon^2),
\label{eq:shadowalpha}
\end{equation}
and Eq.~\eqref{eq:finalqnm} becomes
\begin{equation}
\operatorname{Re}\omega_{\ell n}=\frac{L}{R_{\text{sh}}}+\mathcal{O}(L^{-1}),
\qquad
\operatorname{Im}\omega_{\ell n}=-\left(n+\frac12\right)\lambda_{\text{ph}}+\mathcal{O}(L^{-1}).
\label{eq:shadowqnm}
\end{equation}
Equation~\eqref{eq:shadowqnm} shows that in the eikonal sector the ringdown frequency scale is exactly the inverse shadow scale, while damping is controlled by the photon-sphere instability, in agreement with the geometric-optics correspondence \cite{Cardoso2009,Stefanov2010}.

\section{Strong-lensing correspondence and observables}

To connect directly with the lensing framework of Ref.~\cite{Stefanov2010}, we introduce the standard strong-deflection quantities for a static spherical metric written as
\begin{equation}
ds^2=-A(r)dt^2+B(r)dr^2+C(r)d\Omega^2,
\end{equation}
with, in our case, $A=f$, $B=f^{-1}$ and $C=r^2$. The impact parameter for a null ray with turning point $r_0$ is
\begin{equation}
u(r_0)=\sqrt{\frac{C(r_0)}{A(r_0)}}=\frac{r_0}{\sqrt{f(r_0)}}.
\end{equation}
At the photon sphere, $r_0\to r_{\text{ph}}$, this becomes the critical impact parameter
\begin{equation}
u_m\equiv u(r_{\text{ph}})=\frac{r_{\text{ph}}}{\sqrt{f(r_{\text{ph}})}}=\frac{1}{\Omega_{\text{ph}}}.
\label{eq:umOmega}
\end{equation}

In the strong-deflection expansion of Ref.~\cite{Bozza2002}, as used in Ref.~\cite{Stefanov2010}, the deflection angle for images close to the photon ring is
\begin{equation}
\hat\alpha(\theta)=-\bar a\,\ln\!\left(\frac{\theta D_{OL}}{u_m}-1\right)+\bar b+\mathcal{O}(u-u_m),
\label{eq:deflectionSDL}
\end{equation}
where $D_{OL}$ is the observer--lens distance, and $\bar a,\bar b$ are model-dependent strong-lensing coefficients. The coefficient $\bar a$ is
\begin{equation}
\bar a=\sqrt{\frac{2A_mB_m}{C_m''A_m-C_mA_m''}},
\end{equation}
evaluated at $r_m=r_{\text{ph}}$. For $A=f$, $B=1/f$, $C=r^2$, one gets
\begin{equation}
\bar a=\sqrt{\frac{2}{2f_m-r_m^2f_m''}}.
\label{eq:abarMetric}
\end{equation}
Comparing Eq.~\eqref{eq:abarMetric} with Eq.~\eqref{eq:omegalambda}, we find the exact identity
\begin{equation}
\bar a=\frac{\Omega_{\text{ph}}}{\lambda_{\text{ph}}}.
\label{eq:abarOmegaLambda}
\end{equation}
Hence, both lensing and eikonal ringdown are controlled by the same two geodesic invariants $(\Omega_{\text{ph}},\lambda_{\text{ph}})$.

The directly observable strong-lensing quantities used in Ref.~\cite{Stefanov2010} are
\begin{equation}
\theta_\infty=\frac{u_m}{D_{OL}},
\qquad
s=\theta_\infty\exp\!\left(\frac{\bar b-2\pi}{\bar a}\right),
\qquad
\mathcal{R}=\exp\!\left(\frac{2\pi}{\bar a}\right),
\end{equation}
and the magnitude difference is $r_m=2.5\log_{10}\mathcal{R}$. Using Eqs.~\eqref{eq:umOmega} and \eqref{eq:abarOmegaLambda},
\begin{equation}
\theta_\infty=\frac{1}{D_{OL}\Omega_{\text{ph}}},
\qquad
\mathcal{R}=\exp\!\left(2\pi\frac{\lambda_{\text{ph}}}{\Omega_{\text{ph}}}\right),
\qquad
r_m=\frac{5\pi}{\ln 10}\frac{\lambda_{\text{ph}}}{\Omega_{\text{ph}}}.
\label{eq:lensingInvariants}
\end{equation}

Substituting our small-coupling expansions gives
\begin{equation}
u_m=3\sqrt{3}M\left[1-\frac{\mu\,\varepsilon}{\nu\,3^{\nu}M^{\nu}}\right]+\mathcal{O}(\varepsilon^2),
\qquad
\bar a=1+\frac{\mu(\nu+3)}{2\,3^{\nu+1}M^{\nu}}\,\varepsilon+\mathcal{O}(\varepsilon^2),
\label{eq:umabarExpanded}
\end{equation}
and therefore
\begin{equation}
\theta_\infty=\frac{3\sqrt{3}M}{D_{OL}}\left[1-\frac{\mu\,\varepsilon}{\nu\,3^{\nu}M^{\nu}}\right]+\mathcal{O}(\varepsilon^2),
\qquad
r_m=\frac{5\pi}{\ln 10}\left[1-\frac{\mu(\nu+3)}{2\,3^{\nu+1}M^{\nu}}\,\varepsilon\right]+\mathcal{O}(\varepsilon^2).
\label{eq:thetaRmExpanded}
\end{equation}

Finally, replacing $\Omega_{\text{ph}}$ and $\lambda_{\text{ph}}$ in Eq.~\eqref{eq:masterqnm} by lensing observables yields the explicit model-independent Stefanov-type correspondence (restoring $c$):
\begin{equation}
\operatorname{Re}\omega_{\ell n}=\frac{Lc}{D_{OL}\theta_\infty}+\mathcal{O}(L^{-1}),
\qquad
\operatorname{Im}\omega_{\ell n}=-\left(n+\frac12\right)\frac{c\ln\mathcal{R}}{2\pi D_{OL}\theta_\infty}+\mathcal{O}(L^{-1}).
\label{eq:StefanovMap}
\end{equation}
Equation~\eqref{eq:StefanovMap} provides a direct way to infer ringdown scales from strong-lensing data, and inversely to constrain $(\alpha,\mu,\nu)$ by combining imaging and quasinormal-mode measurements.

\section{Conclusion}

We have derived a fully analytic eikonal sector for quasi-topological regular black holes based on the explicit geometry. The first-order WKB plus $(\varepsilon,1/\ell)$ expansion yields closed quasinormal frequencies with explicit dependence on $(M,\mu,\nu,\alpha)$, providing a complementary analytic counterpart to previous numerical QNM studies. We further showed that the same two null-orbit invariants, $\Omega_{\text{ph}}$ and $\lambda_{\text{ph}}$, govern not only ringdown but also shadow size and strong-deflection lensing observables. This establishes an explicit QNM--shadow--lensing bridge for this metric family, with direct phenomenological use: imaging data $(\theta_\infty,\mathcal{R},r_m)$ and ringdown measurements can be combined to constrain quasi-topological deformation parameters in a unified way.

Our conclusions should be interpreted within the domain of validity of the approximation scheme: first-order WKB, eikonal scaling, and effectively single-barrier potentials. In WKB-unfriendly situations, especially when the effective potential develops a double-well/double-barrier structure, the mode spectrum can contain long-lived branches and the naive geodesic/QNM map can break down or require substantial refinement \cite{KonoplyaStuchlik2017,Konoplya2022Duality,KonoplyaStashko2025Transition}. In parallel, recent rotating regular-black-hole analyses clarify that QNM--shadow correspondence remains powerful when separability and related structural conditions are satisfied \cite{PedrottiVagnozzi2024}.

\section*{Acknowledgments}
The author thanks R. A. Konoplya for the useful discussions. 
The author acknowledges the University of Seville for their support through the Plan-US of aid to Ukraine.

\bibliographystyle{unsrt}
\bibliography{references}

\end{document}